\documentstyle[12pt]{elsart}

\newcommand{\lam}{\lambda}
\newcommand{\eps}{\varepsilon}

\newcommand{\beq}{\begin{equation}}
\newcommand{\eeq}{\end{equation}}
\newcommand{\ba}{\begin{array}}
\newcommand{\ea}{\end{array}}
\newcommand{\beqa}{\begin{eqnarray}}
\newcommand{\eeqa}{\end{eqnarray}}
\newcommand{\bd}[1]{ \mbox{\boldmath $#1$}  }
\begin{document}
\input{epsf.tex}
\begin{frontmatter} 
\title{Semi-Microscopical Description of the Scission 
Configuration in the Cold Fission of  $^{252}$Cf}
\author[nipne]{\c S. Mi\c sicu}
\author[cenbg]{and P. Quentin}

\address[nipne]{National Institute for Physics and Nuclear Engineering,
Bucharest-M\u agurele, Romania}
\address[cenbg]{Centre d'Etudes Nucl\'{e}aires de Bordeaux--Gradignan,
Universit\'{e} Bordeaux I and IN2P3/CNRS, France}

\begin{abstract}
The cold(neutronless) fission of $^{252}$Cf is studied in the frame of a molecular 
model in which the scission configuration is described by two aligned 
fragments interacting by means of Coulomb ( + nuclear) forces. The study is
carried out for different distances between the fragments tips and  
excitation energies. For a given deformation, the fragment's total energy
is computed via the constrained Hartree-Fock + BCS formalism.   
The total excitation energy present in the fragments is supposed 
to contribute only to the fragments deformation and the asymptotic 
value of the kinetic energy is equated to the inter-fragment potential
at scission. These two constraints yield not more than one or two fission 
channels for a fixed tip distance and excitation energy. Discarding those 
fission channels corresponding to a disequilibrated sharing of the excitation
energy between the two fragments, we were able to establish the most likely 
scission configurations for a specified excitation energy. 

\begin{center} PACS numbers 21.60 Gx, 21.60.Jz, 24.75.+i, 25.85.Ca  

\end{center}  
\end{abstract}
\end{frontmatter}    

\section{Introduction}

In last time a renewed interest in the spontaneous fission (sf) of $^{252}$Cf 
arised in connection with  modern experimental techniques, based on large 
Ge detector arrays, which allow a better determination of the mass, charge 
and angular momentum content of the fragments\cite{akop94}. 

A particular attention has been payed to the limiting case of {\em cold
fission}, when no neutrons are emitted and the energy released in the reaction
is converted entirely in the kinetic energy of the fragments. Some features of
this process have been very recently explained satisfactory using cluster like
models\cite{sand98,cron98}. In these models it is assumed that at scission 
the fragments have very compact shapes, close to the ground state and thus they
are carrying very small excitation energy. The scission configuration consists 
of two co-axial fragments with a certain distance $d$ between their tips. 
In the model proposed by G\"onnenwein et al.\cite{goen90}, the cold fission was 
studied by determining the distance $d_{min}$ of the closest approach between 
the two fragments, when the $Q$-value equals the interaction energy. This
model predicted the smallest tip distance (bellow 3 fm) for fragments, with 
mass numbers between 138 and 158 and around the double magic $^{132}$Sn,
emerging in the sf of $^{252}$Cf. Small tip distances were 
interpreted as a sign of cold fission due to the higher interaction energies at 
scission.

In the past the scission-point model succeded also to explain
roughly some basic observables of low-energy fission. Based on the assumption 
of statistical equilibrium among the collective degrees of freedom at the 
scission point, Wilkins et al. \cite{wilk76} calculated the relative 
probabilities of formation of complementary fission fragment pairs from the 
relative potential energies of a system  of two coaxial, quadrupole 
deformed liquid drops, with shell corrections taken into account. The distance
between their tips, the intrinsic excitation energy and collective temperature
were choosen as the free parameters of the model. In this way the general 
features of the distributions of mass, nuclear charge and kinetic energy in the 
fission of various nuclides, ranging from Po to Fm were well reproduced.
Using similar ideas, N\"orenberg\cite{nor72} computed the level schemes, 
equilibrium deformations of the fragments, total energies and charge 
distributions of $^{236}$U, $^{240-242}$Pu using the BCS wave-function in the
description of the ground state.     

In this paper we generalize the static scission-point concept of nuclear 
fission model in such a way that instead of describing the fragments as two 
deformed nearly touching liquid drops with shell corrections taken into account, 
we incorporate the fragments shell structure by means of the self-consistent 
Hartree-Fock(HF) method. For the given binary splitting 
$^{252}$Cf$\rightarrow ^{104}$Mo + $^{148}$Ba we first established the equilibrium
deformations of the two fragments by seeking the HF minimum and next their 
total energy for various deformations is computed by constraining their 
quadrupole moments.
The two fragments are considered as coaxial with distance $d$ between their tips.

One of the basic approximation employed in this paper was that the interaction 
energy at scission is transformed into kinetic energy of the fragments
at infinity. Thus, all the excitation energy present in the fissioning system
is accounted by the deformation energy. This amounts to neglect that 
part of the energy released at the descent from saddle to scission which is
spent on heat. Thus, our study concerns mainly the low-energy domain 
of sf including the limiting case of cold fission
\cite{goen90,san96}. 

By using the above mentioned constraints we were able to deduce the possible 
shapes of the fragments for various tip distances and total excitation
energies $E^*$.
 
\section{Molecular Model of Low Energy Fission}

\subsection{Energy Balance}

In the sf of $^{252}$Cf the fragments are born with a 
certain deformation and will carry a total excitation energy $E^*$, gained 
during the descent from saddle to scission which will be dissipated by means 
of neutron and gamma emission\cite{brosa90}
\beq
Q = V_{sciss} + TKE_{pre} + \sum_{1,2}E_{def}(i)+\sum_{1,2}E_{int}(i) 
\label{release}
\eeq
where $V_{sciss} = V_{coul}+V_{nucl}$ represents the fragments interaction 
energy at scission. For $V_{coul}$ we choose the form corresponding to two 
deformed homogenously charged nuclei with collinear symmetry axes with a distance
$R$ between their centers \cite{cohen62}
\beq
V_{coul} = \sum_{\lam_1=0}^{\infty}\sum_{\lam_2=0}^{\infty}
\frac{3}{{\hat\lam_1}^2({\hat\lam_1^2+2})} 
\frac{3}{{\hat\lam_2}^2({\hat\lam_2^2+2})}
\frac{(2\lam_1+2\lam_2)!}{(2\lam_1)!(2\lam_2)!}x_1^{2\lam_1}x_2^{2\lam_2}
\eeq
The variables $x_1$ and $x_2$ are expressed in terms of the semiaxes $a$
and $b$
\beq
x_{1,2}=\frac{a_{1,2}^2-b_{1,2}^2}{R^2}
\eeq 
The above double series is converging for $|x_1|+|x_2| < 1$ and the final 
result is given, according to \cite{quent69} :
\beqa
V_{coul} &=& \frac{3Z_1Z_2 e^2}{40 R^2}\left\{\frac{1}{x_1^2x_2}
(1 + 11 x_1^2 + 11 x_2^2) + P_x P_y\left [ 
\frac{(1+x_1+x_2)^3}{x_1^3x_2^3}\right.
 \right.  \nonumber \\ 
& &\left.\left. 
\ln (1+x_1+x_2)(1-3x_1-3x_2+12x_1x_2-4x_1^2-4x_2^2)\right ]\right \}
\eeqa
For the attractive nuclear potential we choose the proximity
formula for two nuclei with a finite surface thickness\cite{bloki}
\beq
V_{nucl}=4\pi\bar{R}\gamma\Phi(\zeta)
\eeq
The explanations of the different quantities entering in the above 
formula can be found in \cite{shi86}. 
The prescission kinetic energy $TKE_{pre}$ is taken to be zero, an assumption
which proved to be reasonable for low-energy fission\cite{wilk76}. 
Also, that part of the excitation energy which is transformed into internal 
excitation energy $E_{int}^*$, is neglected. According to calculations based 
on time dependent quantum many-body calculations, the intrinsic excitation 
energy accounts for less than 15\% of the collective energy gain in going from the
saddle to the scission \cite{wali88}. Sch\"utte and Wilets \cite{schut75}
gave also an upper bound for $E_{int}^*$, which is still small compared 
to the total excitation energy $E^*$. 

\subsection{The deformation energy in the frame of the Hartree-Fock + BCS method}

That part of the excitation energy which goes into the deformations of the 
fragments was denoted in eq.(\ref{release}) by $E_{def}$. In the study of
sf properties, $E_{def}$ is taken usually as a sum of the liquid drop 
model (LDM) energy, and the shell and pairing corrections \cite{vandenbosch74}.
In this paper the deformation energy $E_{def}$ of the fissioning system 
at scission is referred to the HF+BCS energy of the two fragments in their 
ground states
\beqa
E_{def} = &E_{HF+BCS}(N_1,Z_1,\beta_1) - E_{HF+BCS}(N_1,Z_1,\beta_1^{g.s.})&
\nonumber\\ 
  +& E_{HF+BCS}(N_2,Z_2,\beta_2) - E_{HF+BCS}(N_2,Z_2,\beta_2^{g.s.})& 
\label{etot}
\eeqa
Obviously this is a more general approach. The LDM, which is based on a 
semiclassical description of the nuclei, supplemented by the shell-effect
corrective energy, is only a poore substitute for a self-consistent calculation.
One of the main advantages of the self-consistent HF+BCS calculation is that
it provides simultanously both the single-particle and semiclassical 
properties of nuclei. The general properties of the Hartree-Fock method
were reviewed in \cite{qu&fl78,rs80}.  

In our study for the HF part of the interaction we choosed the Skyrme interaction
SIII \cite{beiner75}, which succeded to reproduce satisfactory the single-particle
spectra. The difference between the binding energy computed with SIII and the
experimental one appears to be, for a large number of nuclei, $\approx$ 5 MeV
\cite{quent75}. It also produces a fairly well $N - Z$ dependence of the 
binding energy\cite{tajima93}.  The present study envisages nuclei that are not     
in a closed shell configuration. Thus, the level occupations will have a large 
effect on the solution of the HF equations. Usually the HF method is extended to
the Hartree-Fock Bogolyubov (HFB) formalism by using a mixture of different
configurations in place of a single Slater determinant. However, when dealing 
with a Skyrme force which has been simplified such that the bulk properties of
the nucleus are reproduced, one would have to introduce additional parameters 
in order to guarantee that sensible pairing matrix elements are obtained.

Following Vautherin \cite{vauth73} we assign to each orbital $\phi_k$ an 
occupation $n_k=v_k^2$, where $u_k^2 + v_k^2 = 1$, $u_{\bar k} = u_k$ and 
$v_{\bar k} = - v_k$. In terms of the density 
$\rho(\bd{r})=2\sum_k' n_k |\phi_k(\bd{r})|^2$ the HF+BCS total energy, that has
to be minimized reads 
\beq
E_{HF+BCS}=tr\left [ (T + {1\over 2}\cal{V})\rho\right ] +E_p
\label{etotal}
\eeq
where 
\beq
\langle T \rangle = \frac{\hbar^2}{m}\left ( 1 - {1\over A} \right )
{\sum_k}' n_k \int d\bd{r} |\phi_k(\bd{r})|^2
\eeq
is the expectation value of the kinetic energy, ${\cal{V}}=tr(\rho {\tilde{v}})$ 
enters as Hartree-Fock-like potential, $\tilde{v}$ being the antisymmetrized 
effective two-body interaction. The primed sum $\sum'$ denotes a sum over 
all HF orbitals having projections of the total angular momentum $\bd{j}$ on the 
$z$-axis $\Omega_k > 0$. To the total energy we added the pairing
energy
\beq
E_p = - {G\over 4}\left \{ \sum_{k}\left [ n_k(1-n_k)^{1\over 2}\right ]\right\}^2
\eeq 
For BCS-like calculations, the matrix elements of $\tilde{v}$ between HF states 
is taken to be constant
\beq
G = - \int d\bd{r} \int d\bd{r}' \phi_k^*(\bd{r}) \phi_{\bar k}^*(\bd{r}) 
\tilde{v}(\bd{},\bd{r}') \phi_l(\bd{r}) \phi_{\bar l}(\bd{r})   
\eeq 
Varying the normalized single-particle wave functions $\phi_k$ and their 
amplitudes $v_k$ under the additional constraint 
$\lambda_{\tau}\sum_k\left (\delta_{\tau_k,\tau}n_k - N_\tau \right ),
(\tau=p,n)$, ensuring that on the average the system contains the correct number
of neutrons $N$ and protons $Z$, we are lead to the standard HF and BCS equations
\cite{vauth73}. 

The occupations $n_k$ are determined at each step of the HF iterative calculation
using the HF eigenvalues $\eps_k$ and they are employed at the next step to 
construct the HF field. The pairing force constant is
\beq
G_\tau = \frac{G_{0\tau}}{11+N_\tau}~MeV~~~(\tau=p,n)
\eeq  
The constant $G_{0\tau}$ was adjusted  in such a way to obtain the experimental 
pairing gap
\beq
\Delta_{\tau} = G{\sum_{k}}'u_k v_k
\eeq
In the deformed HF calculations one have to optimize the basis which is choosen
to correspond to an axial symmetric deformed harmonic-oscillator with
frequencies $\omega_{\perp}$ and $\omega_z$. Such a basis is characterized by
the deformation parameter $q={\omega_\perp}/{\omega_z}$ and harmonic oscillator
length $b=\sqrt{m\omega_0/\hbar}$, with $\omega_0^3=\omega_\perp^2\omega_z$.
The basis is cut off after $N_{max}$ major shells, where $N_{max}$=10 or 12
for the nuclei emerging in the sf of $^{252}$Cf \cite{floc73a}. 

The next step consists in mapping out the potential energy curves by constraining 
our HF+BCS calculations in which a quadratic constraint ${C \over 2}(Q-Q_0)^2$
is added to the energy functional (\ref{etotal}) \cite{floc73b}. Here $Q_0$ 
is a specified value of the mass quadrupole moment.
In Fig.1 we represent the deformation energy curves of the nuclei $^{104}$Mo
and $^{148}$Ba produced in the sf of $^{252}$Cf.

                         
\section{Distribution of the excitation energy in the fission fragments}

The scope of this section is to seek the configuration of the system at scission
for a fixed excitation energy $E^*$. According to eq.(\ref{release}), the 
interaction energy of two fragments with deformations $\beta_1$ and $\beta_2$
at scission is related to the excitation energy through the relation
\beq
V(\beta_1,\beta_2,d) = Q - E^*
\label{vscis} 
\eeq
where $d$ is the tip distance and enters in the theory as a parameter.
We equate this last quantity with the asymptotic kinetic energy $TKE(\infty)$.
This relation is a conequence of the approximations that we made  earlier,
i.e. we neglected the prescission kinetic energy $TKE_{pre}$ and we forced all
the available excitation energy to be stored into deformation
\beq
E^*(\beta_1,\beta_2) = E_{def}(\beta_1) + E_{def}(\beta_2)
\label{excit}
\eeq 
where $E_{def}$ is computed according to eq.(\ref{etotal}). Thus, for a given 
excitation energy we obtain two non-linear equations, i.e. eqs. (\ref{vscis})
and (\ref{excit}). 

In fig.2 we represented the excitation energy landscape (\ref{excit}),
for the pair ($^{104}$Mo, $^{148}$Ba). The deepest minimum corresponds to the 
prolate-prolate configuration $(\beta_1, \beta_2)$. At this point $E^*=0$ and
fission proceeds by means of only one channel, customarly known as cold fission.
This configuration has deformations $\beta(^{104}$Mo)=0.370 and 
$\beta(^{148}$Ba)=0.270 which are very close to those computed in the frame of 
the finite-range droplet macroscopical model \cite{moeller95}. The non-linear
equations, quoted above, admit this solution only for the tip distance 
$d =$ 2.95 fm, a value very close to the border of 3 fm, allegated by the
T\"ubingen group, bellow which cold fission occurs \cite{cron98}.


When we increase the excitation energy, an infinity of solutions arise according 
to eq.(\ref{excit}). They have to be identified with the geometrical locus of 
points with equal excitation energy. However, the second constraint (\ref{vscis})
is limiting drastically the number of ($\beta_1, \beta_2$) pairs. In Fig. 3  
we give the contour plots of the excitation energy and superposed on them the 
curves relating $\beta_1$ to $\beta_2$ for different tip distances. The 
intersection of such curves with the contour lines of equal excitation energy 
will give the physical solutions to our fission problem, i.e. for certain 
tip distance intervals, one get different scission configurations or channels.



As one observe one get generally two solutions which are located mainly for 
low-excitation energy in the quadrant with $\beta_1, \beta_2 > 0$. While for 
pure cold fission ($E^* = 0$) one get a solution only for one $d$, when 
$E^* > 0$ one get solutions for several values of $d$. Naturally, one may ask 
next if all these solutions are likely to occur. For that one should look at 
the ratio of excitation energies between the two fragments. Calculations based 
on the cascade evaporation model predicted a ratio of the mean excitation 
energies $E^*_2/E^*_1\approx$ 0.5 around the splitting 104/148 when approaching
the limiting case of cold fission \cite{mart84}. According to the same reference,
disproportions in sharing of the excitation energy should be expected only in 
the vicinity of magic numbers. For our study case we are left only with few
possibilities for a given excitation energy, which are listed in Table I for
$E^*$= 0, 2 ,4 and 6 MeV.

In fig.4 we give the fragments density contour lines for a fixed excitation energy, 
namely  $E^*$=2 MeV and different tip distances.

\section{Conclusions}

Based on a molecular model in which the scission configuration has to fulfill
two main energetic constraints, namely that the interaction between the 
fragments is converted totally into asymptotic kinetic energy and that the 
excitation energy of the fissioning system is accounted only by the deformation
energy, we carried constrained HF+BCS calculations at zero temperature for the
nuclei emerging in the low-energy fission reaction. 
For a fixed excitation energy we varied the distance between the tips of the 
fragments. Each case admits not more than two solutions, i.e. to pairs of 
fragments deformations. The criteria which allowed us to select the valid
scission configuration was that of the excitation energy distribution between 
the fragments. We discarded those configurations with a disproportionate
ratio between the excitation energis of the two fragments. A carefull analyse
exquibite roughly two regions of tip distance which can be assigned as 
valid scission configurations. The first one has the starting point at 
$d=2.95$ fm and goes up to 3.8 fm whereas the second is much narrow and is 
centered around $d=5fm$. In this last case that should be less probable 
to occur, one of the fragments is emitted with oblate deformations.     

For excitation energies higher
than those considered in this paper, the properties of the nuclear system
are described by a thermal average, the influence of the shell effects 
becoming thus less important.

A limitation of the present approach is caused by the absence of higher multipole
deformations (octupole, hexadecupole, etc.). As have been shown very recently
\cite{sand98}, the account of hexadecupole deformation provided the explanation 
of a whole region of cold fission for $^{252}$Cf.  

Also we intend to study the fragments angular momentum formation in these 
fragments based on a very recent proposal of some of us \cite{mq99} together
with the evolution of several collective variables during the post-scission
motion.

{\bf Acknowledgements}  One of the authors(\c S.M.) would like to acknowledge 
the financial support from CIES-France. 
He is also very indebted to N.Pilet, I.N.Mikhailov and dr.A.Florescu for
fruitfull discussions. 

\newpage

{\bf Figure legends}
\vskip 2.cm

{\bf Fig.1}
{The deformation energy curves of the nuclei $^{148}$Ba and $^{104}$Mo
computed in the frame of the HF+BCS method with quadratic constrain for the
mass quadrupole}

{\bf Fig.2}
{Three-dimensional plot of the excitation energy $E^*$ for the pair
($^{148}$Ba, $^{104}$Mo) computed in the frame of the HF+BCS method.}

{\bf Fig.3}{Graphical solution of the non-linear equations (\ref{vscis}) and 
(\ref{excit}). The intersection of the solid curve with the contour lines 
provides two solutions in the particular case of the pair
($^{148}$Ba, $^{104}$Mo), with tip distance $d$ = 3.25 fm and total excitation
energy $E^*$ = 2 MeV.}

{\bf Fig.4}{ Fragments density contour lines for excitation energy 
 $E^*$=2 MeV and tip distances $d=$ 2.6, 2.95, 3.10, 3.25}

\begin{table}
\caption{Pairs of fragments deformations $(\beta_1,\beta_2)$ and ratio of 
excitations energies $E^*_2/E^*_1$ for different excitation energies $E^*$
and tip distances $d$.}
\begin{tabular}{c c c c c }                        
$E^*$(MeV) & $d$ (fm)& $\beta_1$& $\beta_2$& $E^*_2/E^*_1$ 
\\
\hline
    0      &  2.95  &  0.270  & 0.370 & - \\
\hline
     2     &  2.60  &  0.313  & 0.492 & 0.94\\
           &  2.65  &  0.325  & 0.463 & 0.30\\
           &        &  0.263  & 0.533 & 260.\\
           &  2.95  &  0.215  & 0.486 & 0.8 \\             
           &        &  0.335  & 0.351 & 0.01\\
           &  3.00  &  0.209  & 0.475 & 0.5\\  
           &        &  0.332  & 0.338 & 0.1\\
           &  3.10  &  0.198  & 0.452 & 0.2\\
           &        &  0.322  & 0.313 & 0.5\\ 
           &  3.15  &  0.194  & 0.438 & 0.2\\
           &        &  0.317  & 0.302 & 1.4\\
           &  3.20  &  0.309  & 0.293 & 1.1\\   
\hline
    4      &  2.60  &  0.342  & 0.520 & 0.70\\
           &  3.10  &  0.182  & 0.530 & 0.91\\
           &        &  0.366  & 0.324 & 0.11\\
           &  3.15  &  0.175  & 0.520 & 0.70\\
           &        &  0.362  & 0.311 & 0.20\\   
           &  3.20  &  0.168  & 0.510 & 0.53\\
           &        &  0.357  & 0.300 & 0.30\\ 
           &  3.25  &  0.352  & 0.286 & 0.41\\
           &        &  0.162  & 0.499 & 0.40\\
           &  3.35  &  0.341  & 0.263 & 0.74\\
           &        &  0.151  & 0.475 & 0.31\\
           &  3.40  &  0.335  & 0.252 & 0.98\\
           &        &  0.146  & 0.463 & 0.14\\
           &  3.85  &  0.175  & 0.265 & 0.71\\
           &  4.95  &  0.335  & -0.212& 0.99\\
           &  5.00  &  0.337  & -0.226& 0.9\\ 

\hline
\end{tabular}
\end{table} 

\begin{table}
\setcounter{table}{0}
\caption{(continued)}  
\begin{tabular}{c c c c c }                        
$E^*$(MeV) & $d$ (fm)& $\beta_1$& $\beta_2$& $E^*_2/E^*_1$ 
\\
\hline
   6       &  2.75  &  0.447  & 0.409 & 0.46\\
           &  3.25  &  0.158  & 0.564 & 0.99\\
           &        &  0.388  & 0.306 & 0.14\\
           &  3.30  &  0.151  & 0.555 & 0.81\\
           &        &  0.385  & 0.293 & 0.21\\
           &  3.35  &  0.143  & 0.546 & 0.65\\
           &        &  0.381  & 0.279 & 0.28\\
           &  3.40  &  0.134  & 0.535 & 0.51\\
           &        &  0.377  & 0.265 & 0.38\\
           &  3.45  &  0.128  & 0.526 & 0.43\\
           &        &  0.373  & 0.253 & 0.48\\
           &  3.50  &  0.121  & 0.516 & 0.34\\
           &        &  0.368  & 0.239 & 0.61\\
           &  3.55  &  0.114  & 0.504 & 0.26\\
           &        &  0.364  & 0.226 & 0.73\\
           &  3.60  &  0.108  & 0.493 & 0.20\\
           &        &  0.360  & 0.213 & 0.87\\
           &  5.00  &  0.373  &-0.213 & 0.49\\
           &  5.05  &  0.374  &-0.227 & 0.46\\
           &  5.15  &  0.371  & -0.250& 0.52\\
           &  5.20  &  0.370  & -0.262& 0.58\\
           &  5.25  &  0.367  & -0.272& 0.65\\
           &  5.30  &  0.364  & -0.282& 0.75\\
           &  5.35  &  0.359  & -0.291& 0.90\\
           &  5.40  &  0.353  & -0.300& 1.1\\
\end{tabular}
\end{table} 
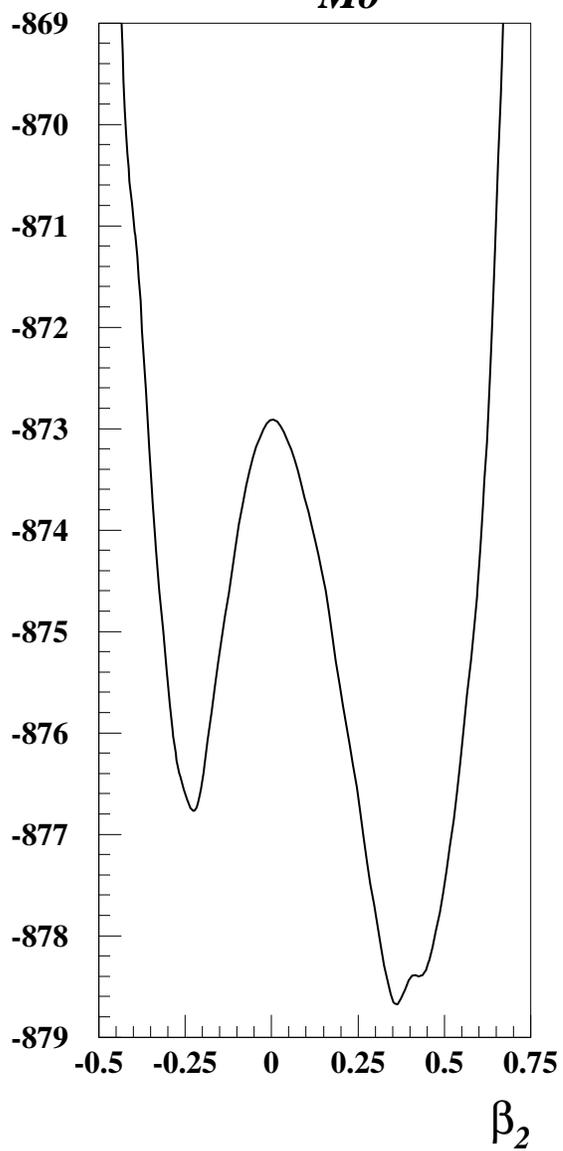
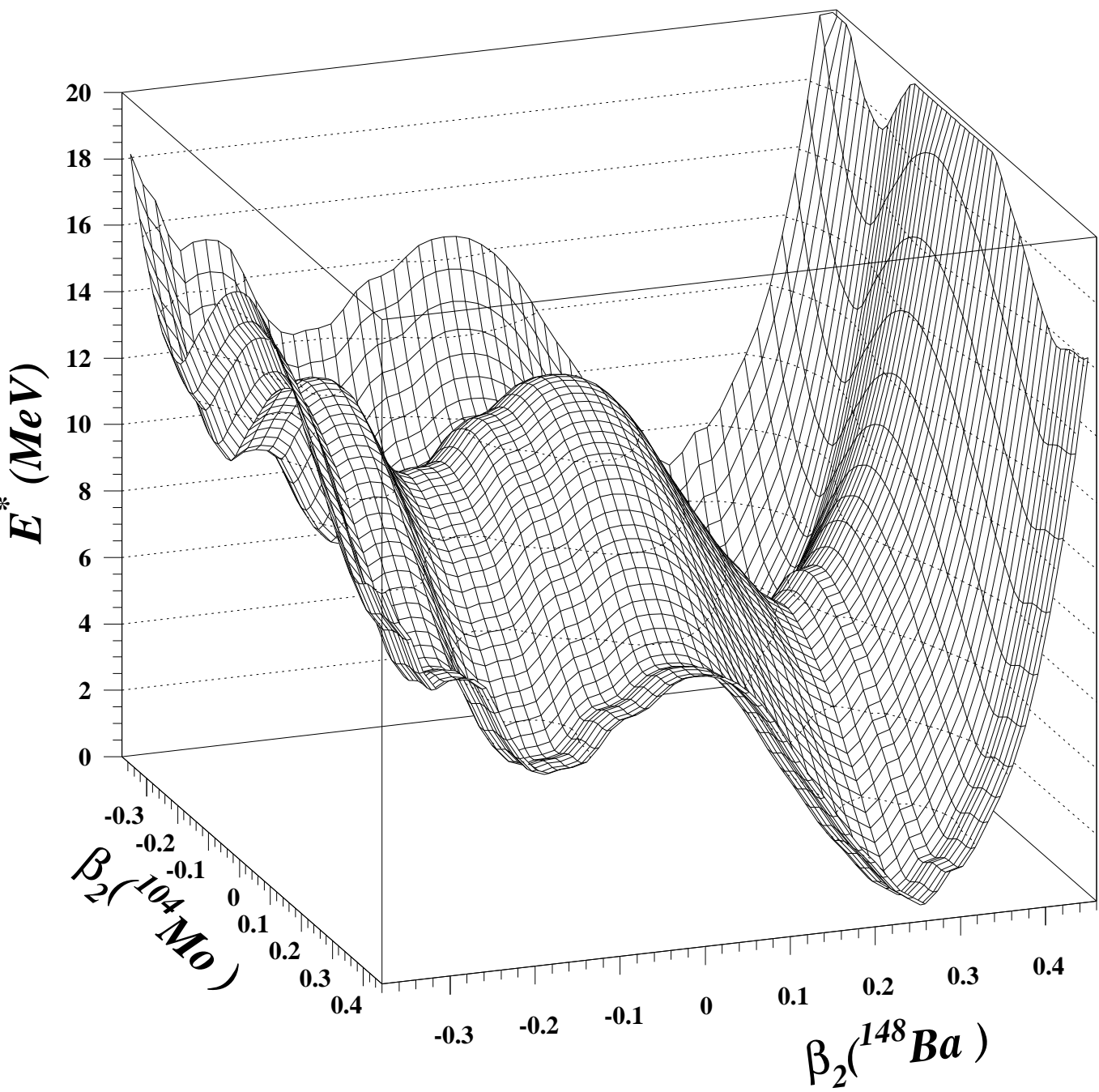
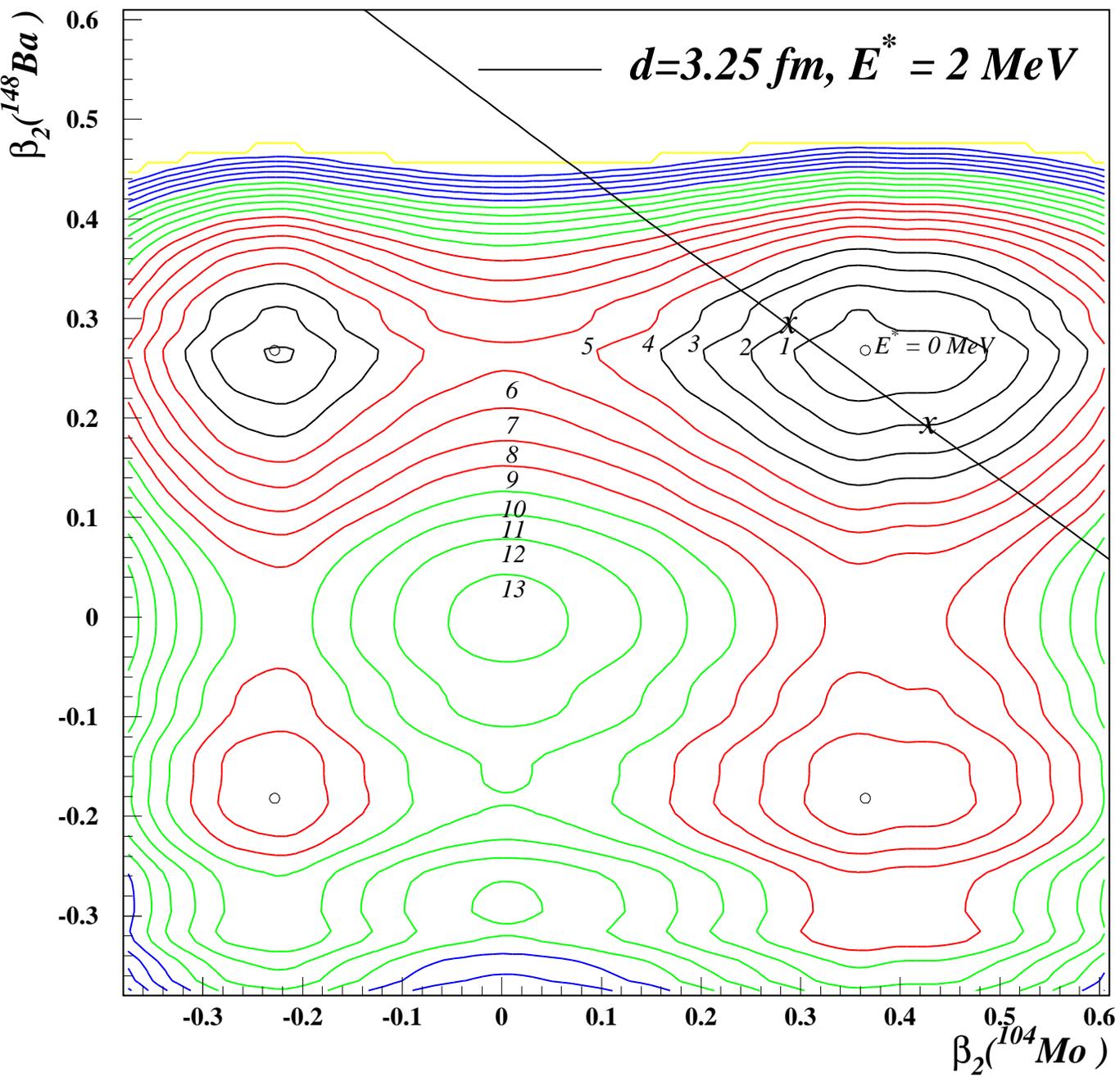
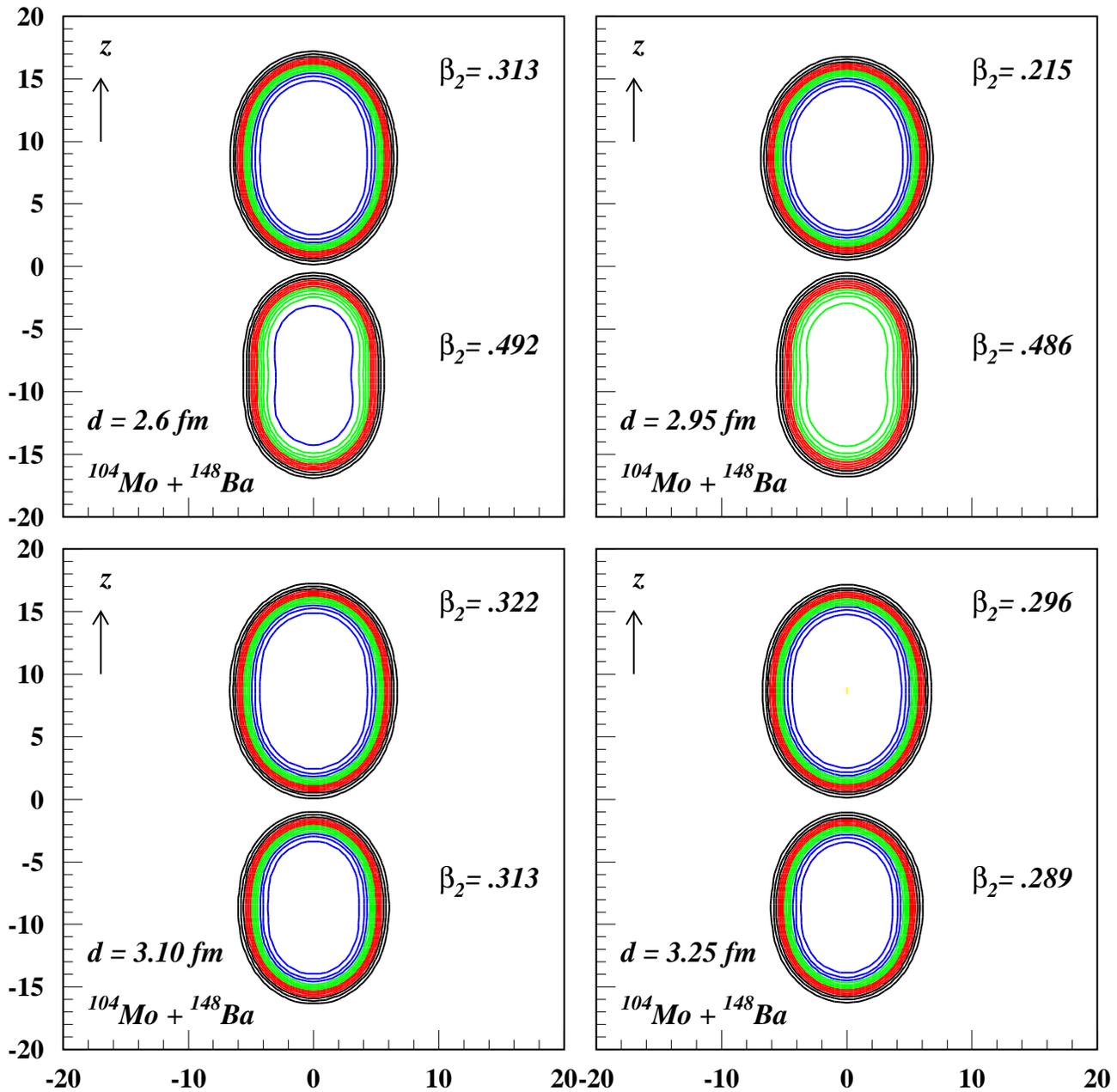
\end{document}